\documentclass[11pt]{article}

\usepackage{amsmath,amssymb,setspace,url}

\usepackage{color}

\usepackage{authblk}
\author[1,2]{Li-Chun Zhang}
\author[2]{Ingvild Johansen}
\author[2]{Ragnhild Nygaard}
\affil[1]{\em \small University of Southampton (email: L.Zhang@soton.ac.uk)}
\affil[2]{\em \small Statistics Norway}
\title{\bf Tests for price indices in a dynamic item universe}
\date{}

\begin{document}

\maketitle

\begin{abstract}  
There is generally a need to deal with quality change and new goods in the consumer price index due to the underlying dynamic item universe. Traditionally axiomatic tests are defined for a fixed universe. We propose five tests explicitly formulated for a dynamic item universe, and motivate them both from the perspectives of a cost-of-goods index and a cost-of-living index. None of the indices satisfies all the tests at the same time, which are currently available for making use of scanner data that comprises the whole item universe. The set of tests provides a rigorous diagnostic for whether an index is completely appropriate in a dynamic item universe, as well as pointing towards the directions of possible remedies. We thus outline a large index family that potentially can satisfy all the tests. 
\end{abstract}



\section{Introduction} \label{introduction}

The failure to take full account of quality change and new goods is one of the important sources of the potential bias of the consumer price index (CPI). See e.g. CPI manual by IWGPS (2004). By convention the term quality change pertains to situations where new products,  models, services etc. are compared to the old items they are deemed to replace, whereas new goods are thought of as wholly new types of items, such as when microwave ovens were first introduced in the market. The cause of the problems is thus the same, namely, the item universe of the CPI is dynamic such that one needs to compare the prices of different sets of items over time.

The challenges are made more urgent in recent years by the greater use of scanner data, where one has access to the average transaction price and quantity of each item (identified by the Global Trade Item Number, GTIN) over the period of data collection. Explicit replacement of the old items by the new ones becomes exceedingly resource demanding, if applied to all the available items. The hedonic methods are often infeasible due to the lack of characteristics data. The research is active at the moment regarding index formulae that are only based on the price and quantity data. See e.g. Chessa et al. (2017), Dal\'{e}n (2017), Diewert and Fox (2017), Zhang et al. (2017). But there is currently a lack of consensus on how to evaluate them. While the choice of index formula may not seem to have a big impact for certain consumption segments, including food (e.g. ABS, 2016; Chessa et al., 2017; Zhang et al., 2017), the choice does matter in many other situations, such as when there is a high item churn rate, and/or where the prices of goods undergo strong decline during their respective life spans. See e.g. Chessa et al. (2017) for some relevant empirical evidence on Men's T-shirt and Television. 

The test approach provides a valuable theoretical framework to index numbers, in addition to the economic and stochastic approaches. However, the traditional axiomatic tests (Fisher, 1922; IWGPS, 2004, Chapter 16) are all defined for a fixed item universe. Diewert (1999) and Balk (2001) outline tests for international comparisons. In this paper, we propose five tests explicitly formulated for a dynamic item universe. The tests will be motivated both from the perspectives of a \textit{cost-of-goods index (COGI)} and a \textit{cost-of-living index (COLI)}. As mentioned in paragraph 16.2 of the CPI manual, ``different price statisticians may have different ideas about which tests are important [...]'' Moreover, there is generally the possibility that a test (or axiom), which otherwise seems attractive, may be in conflict with the principles of another theoretical approach to index numbers. Therefore, the plausibility of a test is clearly strengthened if it can be motivated from both the COGI and COLI perspectives.

The proposed tests will be applied to the indices for scanner data, which have so far received the most attention. These include the Geary-Khamis (GK) index (Geary, 1958; Chessa, 2016), the generalised unit value (GUV) index family (Dal\`{e}n, 2001; De Haan, 2001; Auer, 2014), the weighted geometric means (WGM) index family including the time product dummy (TPD) index (de Haan and Krsinich, 2014), and the GEKS index (Ivancic et al., 2011). It is shown that none of them satisfies all the five tests at the same time. We shall outline a large index family, which includes the GUV index family as a subclass in general and the Fisher index as a special case of bilateral index in a fixed item universe. As will be explained, this family of indices can potentially can satisfy all the tests, thus providing impetus for future research.

Overall, extending the test approach to a dynamic universe has at least three advantages. (i) Formulating explicit tests strengthens the rigour of investigation, because a test result is dichotomous by construction whereas an empirical comparison may be inclusive on its own. (ii) It can help to clarify certain conceptual ambiguities. For instance, transitivity can obviously prevent chain drifting in a fixed universe. However, as we shall discuss in Section 2.3, a rigorous definition of transitivity is impossible in a dynamic universe. The proposed tests can be help to articulate the intuition against chain drifting and to avoid an index method that suffers from chain drifting. 
(iii) Where an index fails a test, it points to the direction in which the index can possibly be improved. For instance, it will be shown that the GK or GUV index can fail the responsiveness test T5 in a number of situations. 

The rest of the paper is organised as follows. In Section \ref{test}, the necessary notations and concepts are introduced, following which the tests are defined and explained, including a discussion why some other seemingly obvious tests are not included in the proposal. In Section \ref{index} the tests are applied to the indices mentioned above. Finally, an index family is outlined in Section \ref{outline} which potentially can satisfy all the tests, and a few concluding remarks given in Section \ref{remark}.

\section{Tests for a dynamic item universe} \label{test}

\subsection{Notations}

Let us introduce formally the notations and terms that are necessary for a dynamic item universe. Consider a given \emph{item universe} in period $t$, denoted by $U_t = \{ 1, 2, ..., N_t\}$, which constitutes a subset (or sector) of the entire CPI item universe. For instance, $U_t$ may refer to all food and non-alcoholic beverages sold at supermarkets, or all personal computers sold at electro warehouses, etc. Notice that generally we allow the items to be specific for an outlet or chain. For example, the same mobile phone model sold at two different outlets can be treated as two different items. 
Assume that one has available the unit value price $p_i^t$ and the transaction total $v_i^t$, for each $i\in U_t$, where $v_i^t = p_i^t q_i^t$ and $q_i^t$ denotes the transaction quantity of the item. Notice that the unit value price refers to the average transaction price of an item in the period $t$, where the actual transaction price may change over the period. Denote by $q(U_t) = \{ q_i^t; i\in U_t\}$ the set quantities and by $p(U_t) = \{ p_i^t; i\in U_t\}$ the set of prices. Denote all the data from period $t$ by
\[
D_t = D(U_t) = \{ p(U_t), q(U_t); U_t\}.
\] 
 
Denote by $(U_0, U_t)$ the \emph{comparison universe}, for which we seek a price index from $0$ to $t$, denoted by $P^{0,t}$. We refer to $0$ as the \emph{base} period and $t$ the \emph{statistical} (or \emph{current}) period. Denote by $U_{R(0,t)} = \{ U_r; r\in R\}$ the \emph{reference universe}, where $R$ is the set of reference periods involved, and the notation $U_{R(0,t)}$ emphasises the dependence of $R$ on the comparison universe -- we may suppress $(0,t)$ whenever such an emphasis is unnecessary. We consider price indices that are functions of the data $D_R = D(U_R) = \{ D_t; t\in R\}$, which can be given as  
\begin{equation}
P^{0,t} = f(D_R). \label{scope}
\end{equation}
Two choices of the reference universe $U_R$ are most immediate, i.e.
\begin{gather*}
R_B = \{ 0, t\} \quad\text{and}\quad U_{R_B} = U_0 \cup U_t, \\
R_M = \{ 0, 1, ..., t\} \quad\text{and}\quad U_{R_M} = U_0 \cup U_1 \cup \cdots \cup U_t, 
\end{gather*} 
where $R_B$ implies direct comparison between $0$ and $t$, e.g. December 2014 (0) and July 2015 ($t$), and $R_M$ implies all the periods from $0$ to $t$, e.g. December 2014 to July 2015. An index based on the reference universe $U_{R_B}$ is referred to as a \emph{bilateral} index; it is \emph{multilateral} otherwise. We notice that it is possible to use another reference set than $R_M$ for a multilateral index, such as a rolling window of fixed length counting backwards from $t$. However, the distinction is not important for the discussion in this paper, so we shall adopt the convention of multilateral reference universe $U_{R_M}$ as specified above, unless otherwise noticed.

\subsection{The tests}

Let $U_{0t} = U_0 \cap U_t$ be the \emph{persistent} item universe at $0$ and $t$, and $U_{t\setminus 0} = U_t\setminus U_0$ the set of \emph{birth} items and $U_{0\setminus t} = U_0\setminus U_t$ that of the \emph{death} items. The item universe is \emph{fixed} if $U_t = U_0 = U_{0t}$ and $U_{t\setminus 0} = U_{0\setminus t} = \emptyset$, for any $t\neq 0$; it is \emph{dynamic} otherwise. Below we formulate five tests  for a dynamic item universe, and motivate them from both the COGI and COLI perspectives. Notice that in the case of $t=1$, both the tests T1 and T2 reduce to well-known tests for a fixed item universe. However, under the formulation here, this represents the special case where a dynamic universe may sometimes return to the same state of affairs, in which respect we require that a dynamic-universe index should not then have counter-intuitive properties.

\paragraph{\em Identity test (T1)} If $U_0 = U_t$ and $p_i^0 \equiv p_i^t$ for all $i\in U_0$, then $P^{0,t} =1$.

\medskip 
Since the item universe is the same at $0$ and $t$, so must be the items eligible for a COGI, when the comparison universe is $(U_0, U_t)$. Thus, despite the changes of item universe which take place between the two time points, where $t>1$, the identity test can be motivated for a COGI. Now that all the prices are the same at $0$ and $t$, a basket-of-goods index must necessarily be 1, regardless of how the reference quantities of the goods are calculated. Therefore, a COGI should satisfy the identity test. Next, consider a COLI. Let $V^{0,t} = \sum_{i\in U_t} q_i^t p_i^t/\sum_{i\in U_0} q_i^0 p_i^0$ be the ratio of total expenditures. Under the stipulated setting, it is obviously possible to maintain the same utility without changing the total expenditure. Thus, insofar as $V^{0,t} \neq 1$, all the change in expenditure must be attributed to the change in utility but \emph{not} prices, under the assumption of rational consumer behaviour. A COLI should therefore be equal to 1. 

Despite the universe varies at other points than $0$ and $t$, an index that satisfies the identity test is said \textit{not to drift} in this situation, which can be used to examine e.g. a multilateral index. We notice that chain drift is often contrasted with transitivity. However, as will be discussed in Section \ref{transitivity}, we find it difficult to define a transitivity test for a dynamic item universe. For the moment the identity test T1 is the only test we have with respect to chain drift.

\paragraph{\em Fixed basket test (T2)} If $U_0 = U_t$ and $q_i^0 \equiv q_i^t$ for all $i\in U_0$, then $P^{0,t} = V^{0,t}$.

\medskip 
The test is obvious for a COGI. It can easily be satisfied by any bilateral COGI. Otherwise, in a dynamic universe, one may have $q_i^r \neq q_i^0$, for $0 < r< t$, so that the item reference quantity can differ from $q_i^0 = q_i^t$, and a fixed basket-of-goods index may not be equal to $V^{0,t}$. It follows that, in order for a COGI to satisfy T2, one may need to avoid the use of multilateral indices. The test T2 is readily motivated for a COLI. Given the quantities $q(U_t)$ are actually the same as $q(U_0)$, no utility adjustment of $V^{0,t}$ is needed, and a COLI at the observed utility is equal to $V^{0,t}$.

\paragraph{\em Upper bound test (T3)} If $U_0 \subseteq U_t$, and $p_i^t \leq p_i^0$ for all $i\in U_0$, then $P^{0,t} \leq 1$.

\medskip 
That is, the item universe may be constant if $U_0 = U_t$ or strictly expanding if $U_0 \subset U_t$, and the price of each persistent item is the same or reduced, i.e. $p_i^t \leq p_i^0$ for all $i\in U_{0t} = U_0$. To motivate the test, consider the following. Firstly, suppose substitution does not occur, in which case $q_i^t = q_i^0$ for all $i\in U_0$ and $q_i^t = 0$ for all $i\in U_{t\setminus 0}$, even if $U_{t\setminus 0}$ is nonempty. The actual comparison universe reduces then to $U_{0t}$, so that the test T2 applies, yielding $P^{0,t} = V^{0,t} \leq 1$ under the stipulated setting. Next, suppose substitution occurs only among the persistent items, i.e. $q_i^t = 0$ for $i\in U_{t\setminus 0}$ and $q_i^t \neq q_i^0$ for some $i\in U_{0t}$. Substitution can be accounted for from the perspective of COLI. Given the actual $\{ q_i^t; i\in U_{0t}\}$ and the corresponding utility at $t$, it cannot cost less for the same $\{ q_i^t; i\in U_{0t}\}$ at $0$ since $p_i^t \leq p_i^0$ for all $i\in U_{0t}$. It follows that a COLI must be less or equal to 1. Finally, suppose substitution also involves the items in $U_{t\setminus 0}$. Let $\{ \tilde{q}_i^t; i\in U_{0t}\}$ be a hypothetical set of persistent items that would have yielded the same utility as the actual $\{ q_i^t; i\in U_t\}$. Owing to rational behaviour, the expenditure of $\{ \tilde{q}_i^t; i\in U_{0t}\}$ at $t$ cannot be less than the actual expenditure of $\{ q_i^t; i\in U_t\}$; whereas the expenditure of $\{ \tilde{q}_i^t; i\in U_{0t}\}$ at $0$ cannot be less than that at $t$. It follows again that a COLI must be less or equal to 1.

\paragraph{\em Lower bound test (T4)} If $U_t \subseteq U_0$, and $p_i^t \geq p_i^0$ for all $i\in U_t$, then $P^{0,t} \geq 1$.

\medskip 
That is, the item universe may be constant or strictly shrinking, and the price of each persistent item is the same or increased. Firstly, suppose substitution does not occur, in which case $q_i^t = q_i^0$ for all $i\in U_{0t}$ and $q_i^t = 0$ for all $i\in U_{0\setminus t}$. Then, the comparison universe reduces to $U_{0t}$, and the test T2 applies, yielding $P^{0,t} = V^{0,t} \geq 1$ under the stipulated setting. Next, suppose substitution occurs only among the persistent items. Given any actual $\{ q_i^t; i\in U_{0t}\}$ and the corresponding utility at $t$, it cannot cost more for the same $\{ q_i^t; i\in U_{0t}\}$ at $0$ since $p_i^0 \leq p_i^t$ for all $i\in U_{0t}$. It follows that a COLI must be greater or equal to 1. Finally, suppose substitution also involves the items in $U_{0\setminus t}$. Let $\{ \tilde{q}_i^0; i\in U_{0t}\}$ be a hypothetical set of persistent units that would have yielded the same utility as the actual $\{ q_i^0; i\in U_0\}$. The expenditure of $\{ \tilde{q}_i^0; i\in U_{0t}\}$ at $0$ cannot be less than the actual expenditure of $\{ q_i^0; i\in U_0\}$; whereas the expenditure of $\{ \tilde{q}_i^0; i\in U_{0t}\}$ at $t$ cannot be less than that at $0$. It follows again that a COLI must be greater or equal to 1.

Under the setting of test T3, there exists a clear downwards trend of the prices of persistent items. We should have $P^{0,t} \leq 1$ even if this leads to an increase of expenditure, i.e. $V^{0,t} >1$. Under the setting of test T4, there exists a clear upwards trend of the prices of persistent items. We should have $P^{0,t} \geq 1$ even if the price increase causes the expenditure to drop, i.e. $V^{0,t} <1$. 

It is possible to formulate two somewhat sharper versions of the tests T3 and T4, respectively, according to which $P^{0,t}$ can possibly deviate from 1 in a particular direction depending on whether the item universe is expanding or shrinking, when all the prices of persistent items remain the same. These are thus clearly the implications of the fact that the item universe is dynamic.  

\paragraph{\em Test t3} If $U_0 \subset U_t$ and $p_i^0 = p_i^t$ for all $i\in U_0$, then $P^{0,t} \leq 1$.  

\paragraph{\em Test t4}  If $U_t \subset U_0$ and $p_i^0 = p_i^t$ for all $i\in U_0$, then $P^{0,t} \geq 1$.

\paragraph{\em Responsiveness test (T5)} For $U_0 \neq U_t$,  $P^{0,t}$ should not always reduce to $f(D_{0t})$, where $D_{0t} = D(U_{0t})$.

\medskip 
That is, one should not always be able to reduce a COGI to a fixed-basket index, where the basket items only consists of the persistent items. This is necessary for any COGI that in principle can be applied to a dynamic item universe. Whereas one should not always be able to reduce a COLI to $f(D_{0t})$, since it must allow for substitution that involves the birth and death items.

One can formulate a sharper version of the test T5, where $p_i^0 = p_i^t$ for all $i\in U_{0t}$. According to T1, we have then $P^{0,t}(D_{0t}) = 1$, which is the price index of the persistent item universe $U_{0t}$. Any $P^{0,t}$ that is always equal to 1, regardless of $D(U_{t\setminus 0})$ or $D(U_{0\setminus t})$, is not responsive. 

\paragraph{\em Test t5} For $U_0 \neq U_t$, if $p_i^0 = p_i^t$ for all $i\in U_{0t}$, then $P^{0,t}$ cannot always be equal to 1, regardless of $D(U_{t\setminus 0})$ and $D(U_{0\setminus t})$.

\subsection{Discussion} \label{transitivity}

The proposed tests are certainly not the only ones possible. However, we have not included any other tests here for several reasons. Firstly, it would have made little difference to include a test that is easily satisfied, an example of which is the time reversal test. Next, some tests seem no longer relevant given the birth and death items. The quantity reversal and price reversal tests are two such examples. Moreover, we have excluded some familiar tests and only retained a sharper version of them. An example is the proportionality test. Since the proportionality test implies the identity test T1, the latter is sharper than the former, in the same sense that the test t3 is sharper than T3. Finally, there may be additional concerns which make a test difficult to formulate. An example is the transitivity test, as we discuss below. 

Conceptually, an index is transitive if $P^{0,t} = P^{0,r} P^{r,t}$ for any $r\neq 0, t$, provided all the three indices in the form of \eqref{scope} are calculated in the \emph{same} way, which generally involves three different reference universes $U_{R(0,t)}$, $U_{R(0,r)}$ and $U_{R(r,t)}$ when the item universe is dynamic. Now, a motivation for transitivity is to prevent chain drifting, when chaining is used to alleviate the difficulty one would encounter in making direct price comparisons between $U_0$ and $U_t$, where $U_{t\setminus 0}$ and $U_{0\setminus t}$ may be non-negligible in size compared to $U_{0t}$. However, in order to verify whether or not chain drifting is the case, one must compare the chained index to the direct index that could have been calculated between $0$ and $t$. Thus, one cannot avoid running into the same difficulty that has motivated the chaining in the first place. To push the difficulty to the logical  extreme, suppose $U_0 \cap U_t = \emptyset$, i.e. the item universe is completely renewed. What are the conditions of non-drifting, or transitivity, in this case?

From a more pragmatic point of view, the GEKS index has been adapted for temporal price comparisons (Ivancic et al., 2011), as a means to achieve transitivity, provided $P^{0,r}$ are well-defined and time reversible for any $0< r\leq t$. However, international comparisons have a fixed reference set of countries (or regions), and are symmetric in the sense that any two countries are eligible for comparison. The temporal extension has a direction and is ever-changing. Regarding the direction of time, it seems counter-intuitive to require $P^{0,r} P^{r,t} = P^{0,t}$, for an arbitrarily chosen period $r$, where $r<0$ or $r>t$. Regarding the changing reference set, the GEKS index $P^{0,t}$ calculated at $t$ will generally differ to $P^{0,t}$ calculated at $t'$, for $t' > t$. It follows from both accounts that in reality the disseminated GEKS index is nevertheless intransitive -- see Section \ref{geks} for details.

\section{Application} \label{index}

The test results are summarised in Table \ref{tab-tests}. We show that the constant-value adjustment of the GK index, which is necessary in the context of international comparisons involving different currencies, may lead to nonresponsiveness in a dynamic item universe. Dropping the constant-value adjustment yields the modified GK (MGK) index, which is a member of the GUV index family. We consider also the WGM index family, which includes the TPD index as a special case in a dynamic universe and the T\"{o}rnqvist index as a special case in a fixed universe. Finally, we discuss the GEKS index, which can be based on the bilateral MGK, GUV or WGM index. 

\begin{table}[ht]
\begin{center}
\caption{Summary of test results}
\begin{tabular}{l|c|c|c|c|c}\hline \hline
 & Identity & Fixed-basket & Upper-bound & Lower-bound & Responsiveness\\ \hline
GUV & Yes if $R_B$ & Yes & Yes & Yes & No in Setting \\ 
(MGK) & No if $R_M$ & & & & of t3 or t4 \\ \hline
WGM & Yes if $R_B$ & No & Yes & Yes & No in Setting \\ 
& No if $R_M$ & & & & of t3 or t4 \\ \hline
GEKS & No & No & No & No & No if $(U_0, U_1)$ \\ \hline \hline
\end{tabular} \label{tab-tests}
\end{center} \end{table}

\subsection{The GK index}

Deflating a constant-value reference-price quantity index yields the GK index:
\begin{align}
& P_{GK}^{0,t} = V^{0,t}/Q^{0,t} \qquad\text{and}\qquad 
Q^{0,t}  = \sum_{i\in U_t} p_i q_i^t / \sum_{i\in U_0} p_i q_i^0, \label{GK}\\
& p_i = \sum_{r\in R_i} \frac{p_i^r}{P^{0,r}} q_i^r/\sum_{r\in R_i} q_i^r, \label{chessa}
\end{align}
where the observed price $p_i^r$ is adjusted to a constant-value price by $P^{0,r}$, and $R_i$ contains the periods at which the item $i$ is in the market. When $U_0 \neq U_t$, the bilateral GK index with \eqref{chessa} fails the responsiveness test T5, provided $q_i^r = \delta_i^r$ where $\delta_i^r =1$ if $i\in U_r$ and 0 otherwise, such as when the items are rental objects. Let $R = R_B = \{0, t\}$ for a bilateral index. Let $V_0 = \sum_{U_0} p_i^0 = V_{0t}^0 + V_{0\setminus t}^0$ over $U_{0t}$ and $U_0\setminus U_t$, and $V_t = \sum_{U_t} p_i^t = V_{0t}^t + V_{t\setminus 0}^t$. We have 
\[
P^{0,t} = \frac{V^{0,t}}{Q^{0,t}} = \Big( \frac{V_{0t}^t + V_{t\setminus 0}^t}{V_{0t}^0 + V_{0\setminus t}^0} \Big) / 
\Big( \frac{\sum_{i\in U_t} p_i q_i^t}{\sum_{i\in U_0} p_i q_i^0} \Big) 
 = \Big( \frac{V_{0t}^t + V_{t\setminus 0}^t}{V_{0t}^0 + V_{0\setminus t}^0} \Big) /
 \Big( \frac{\frac{V_{0t}^0}{2} + \frac{V_{0t}^t}{2 P^{0,t}} + \frac{V_{t\setminus 0}^t}{P^{0,t}}}{\frac{V_{0t}^t}{2 P^{0,t}} 
 + \frac{V_{0t}^0}{2} + V_{0\setminus t}^0} \Big),
\]
where $p_i = (\delta_i^0 p_i^0 + \delta_i^t p_i^t/P^{0,t})/(\delta_i^0 + \delta_i^t)$. Successively rearranging the expression yields
\begin{gather*}
P^{0,t} = V_{0t}^t/V_{0t}^0,
\end{gather*}
which depends only on the persistent item universe $U_{0t}$ in this situation.

The constant-value adjustment in the original GK index seems debatable for the CPI, which is calculated based on a single currency. For example, applying the constant-value adjustment to the Laspeyres index would have yielded its squared root as the adjusted Laspeyres index, i.e.
\[
P^{0,t} = \frac{\sum_{i\in U} q_i^0 p_i^t/P^{0,t}}{\sum_{i\in U} q_i^0 p_i^0} \quad\Rightarrow\quad
P^{0,t} = \Big( \frac{\sum_{i\in U} q_i^0 p_i^t}{\sum_{i\in U} q_i^0 p_i^0} \Big)^{\frac{1}{2}}.
\]
A possible remedy is to drop the constant-value adjustment in \eqref{chessa} and use
\begin{equation}
p_i = \sum_{r\in R_i} p_i^r q_i^r / \sum_{r\in R_i} q_i^r. \label{lehr}
\end{equation}
This yields the Lehr index (Lehr, 1885, p.39) as a bilateral index in a fixed universe, and a \emph{modified} GK (MGK) index by \eqref{GK} in a dynamic universe. With $q_i^t = \delta_i^t$, we now have
\[
P^{0,t} = \Big( \frac{V_{0t}^0 + V_{0t}^t + 2 V_{t\setminus 0}^t}{V_{0t}^0 + V_{0t}^t  + 2V_{0\setminus t}^0} \cdot V^{0,t} \cdot 
\frac{V_{0t}^0 + V_{0t}^t  + 2V_{0\setminus t}^0}{V_{0t}^0 + V_{0t}^t + 2 V_{t\setminus 0}^t} \Big)^{\frac{1}{2}}
= \sqrt{V^{0,t}},
\]
which neither reduces to the persistent universe $U_{0t}$ nor equals to the value index $V^{0,t}$. 

The MGK index \eqref{GK} and \eqref{lehr} does not satisfy the identity test T1 except when $R(0,t) = R_B$, in which case $p_i = p_i^0 = p_i^t$. Next, it satisfies the fixed-basket test T2 since $Q^{0,t} = 1$. Thirdly, it satisfies the upper bound tests T3 and t3, provided $p_i^t \leq p_i \leq p_i^0$ by \eqref{lehr}, such that 
\[
Q^{0,t} = \frac{\sum_{j\in U_0} p_j q_j^t + \sum_{j\in U_{t\setminus 0}} p_j^t q_j^t}{\sum_{j\in U_0} p_j q_j^0} \geq \frac{\sum_{j\in U_0} p_j^t q_j^t + \sum_{j\in U_{t\setminus 0}} p_j^t q_j^t}{\sum_{j\in U_0} p_j^0 q_j^0} = V^{0,t}.
\]
Moreover, it satisfies the lower bound tests T4 and t3, since we have then $Q^{0,t} \leq V^{0,t}$. However, it fails the responsiveness test t5 in the setting of tests t3 or t4. In the setting of test t3, we have $p_i = p_i^0 = p_i^t$ for $i\in U_0$ and $p_i = p_i^t$ for $i\in U_{t\setminus 0}$ provided $R_B$, such that
\[
Q^{0,t} = \frac{\sum_{i\in U_t} q_i^t p_i}{\sum_{i\in U_0} q_i^0 p_i} = \frac{\sum_{i\in U_t} q_i^t p_i^t}{\sum_{i\in U_0} q_i^0 p_i^0} = V^{0,t} \qquad\text{and}\qquad P_{GK}^{0,t} = 1,
\]
regardless of $D(U_{t\setminus 0})$. Similarly, in the setting of test t4, we have $P_{GK}^{0,t} = 1$ regardless of $D(U_{0\setminus t})$.

\subsection{The GUV index family}

Replacing the accompanying formula \eqref{chessa} by any suitable $p_i$ yields a family of GUV indices:
\begin{equation}
P_{GUV}^{0,t} = \frac{\sum_{i\in U_t} p_i^t q_i^t / \sum_{i\in U_t} p_i q_i^t}{\sum_{i\in U_0} p_i^0 q_i^0 / \sum_{i\in U_0} p_i q_i^0} 
= V^{0,t}/Q_{RP}, \label{GUV} 
\end{equation}
where $Q_{RP} = \sum_{i\in U_t} p_i q_i^t / \sum_{i\in U_0} p_i q_i^0$ can be formulated as a reference-price (RP) quantity index. Auer (2014) emphasises the interpretation of $p_i$ as an adjustment factor which transforms the transaction quantities $(q_i^0, q_i^t)$ into the ``intrinsic-worth units'' $(\tilde{q}_i^0, \tilde{q}_i^t)$, where $\tilde{q}_i^r = p_i q_i^r$ for $r = 0, t$. What then matters to the resulting index is only the relevant ratios $p_i/p_j$ for any $i\neq j$. As pointed out by Dal\'{e}n (2001), if hedonic regression is used, the factor $p_i$ could be determined based on the difference in characteristics between the item and the numeraire (or chosen reference item). In this way the GUV index family can incorporate the hedonic approach. 

The test results for $P_{GUV}^{0,t}$ are similar to the MGK index above, which is a special case of the GUV index \eqref{GUV}. It does not satisfy the identity test T1 except when $R(0,t) = R_B$. It satisfies obviously the fixed-basket test T2. It satisfies the upper bound tests T3 and t3, provided $p_i^t \leq p_i \leq p_i^0$ for $i\in U_R$. Similarly, it satisfies the lower bound tests T4 and t4 provided $p_i^0 \leq p_i \leq p_i^t$ for $i\in U_R$. However, it fails the responsiveness test t5 in the settings of tests t3 and t4, provided the tests t3 and t4 are satisfied by the choices of $p_i$ for $i\in U_R$.

\subsection{The WGM index family} \label{wgm}

A WGM index does not have a direct connection to the expenditure ratio $V^{0,t}$ in general. Like the GUV index, it employs a reference term $p_j$, for $j\in U_R$, and is given by
\begin{equation}
P_{WGM}^{0,t} = \frac{\prod_{i\in U_t} (p_i^t/p_i)^{w_i^t}}{\prod_{i\in U_0} (p_i^0/p_i)^{w_i^0}} 
= \Big( \frac{\prod_{i\in U_t} (p_i^t)^{w_i^t}}{\prod_{i\in U_0} (p_i^0)^{w_i^0}} \Big) / 
\Big( \frac{\prod_{j\in U_t} p_j^{w_j^t}}{\prod_{j\in U_0} p_j^{w_j^0}} \Big), \label{WGM}
\end{equation} 
with the weights $\sum_{i\in U_t} w_i^t = 1$ and $\sum_{i\in U_0} w_i^0 = 1$, and $p_j$ a reference price of $j\in U_R$. When $R = \{ 0, t\}$ and $U = U_0 = U_t$, setting $w_j^0 = w_j^t = \frac{1}{2} (q_j^0 p_j^0/\sum_{i\in U} q_i^0 p_i^0 + q_j^t p_j^t/\sum_{i\in U} q_i^t p_i^t)$ reduces \eqref{WGM} to the T\"{o}rnqvist index. The TPD index (de Haan and Krsinich, 2014) is given by
\[
p_j = \prod_{r\in R_j} \Big( \frac{p_j^r}{P^{0,r}} \Big)^{\frac{w_j^r}{\sum_{b\in R_j} w_j^b}} \qquad\text{and}\qquad
w_j^r = \frac{q_j^r p_j^r}{\sum_{i\in U_r} q_i^r p_i^r}.
\]
Fattore (2010) considers the axiomatic properties of the geo-logarithmic family (GLF) of indices. The GLF index is a special case of the WGM index, with fixed universe $U_0 = U_t$ and the same weights at both periods $0$ and $t$, i.e. $w_i = w_i^0 = w_i^t$, where $w_i$ can depend on data at both $0$ and $t$.

The WGM index \eqref{WGM} does not satisfy the identity test T1 except when $R(0,t) = R_B$, since otherwise one can not ensure $p_j = p_j^0 = p_j^t$ in a dynamic universe. It generally does not satisfy the fixed-basket test T2 due to the lack of direct connection to $V^{0,t}$. It satisfies the upper bound test T3, provided $p_i^t \leq p_i \leq p_i^0$, such that
\[
P_{WGM}^{0,t} = \Big( \prod_{i\in U_t} (p_i^t/p_i)^{w_i^t} \Big) \Big( \prod_{i\in U_0} (p_i/p_i^0)^{w_i^0} \Big) \leq 1.
\]
Similarly, it satisfies the lower bound test T3, provided $p_i^0 \leq p_i \leq p_i^t$. Under the settings of tests t3 and t4, we have $P_{WGM}^{0,t} = P_{WGM}^{0,t}(D_{0t}) = 1$, provided $p_j = p_j^0 = p_j^t$, such that it satisfies these tests while failing the responsiveness test t5 at the same time.

\subsection{The GEKS index} \label{geks}

Provided $R(0,t) = R_M$ and $t\geq 2$, the GEKS index from $0$ to $r$, for $0< r\leq t$, is given by
\begin{gather}
P_{GEKS}^{0,r} = \Big( \prod_{s=0}^t P^{0,s} P^{s,r} \Big)^{\frac{1}{t+1}} 
= \Big( (P^{0,r})^2 \prod_{s\neq 0,r} P^{0,s} P^{s,r} \Big)^{\frac{1}{t+1}}.  \label{GEKS}
\end{gather}
For any $r< t$, it involves indirect comparisons via the periods outside $\{ 0, ..., r\}$. For example, if $t=2$ and $r=1$, we have $P^{0,1} = \left( (P^{0,1})^2 P^{0,2} P^{2,1} \right)^{\frac{1}{3}}$ where both $P^{0,2}$ and $P^{2,1}$ are only available in period 2 but not period 1. Therefore, in practice, the disseminated GEKS index, denoted by $\hat{P}_{GEKS}^{0,r}$ is always the one with $r=t$ in \eqref{GEKS}. It is intransitive since, for any $0< r <t$,
\begin{align*}
\hat{P}_{GEKS}^{0,t} & = \Big( (P^{0,t})^2 \prod_{0<s<t} P^{0,s} P^{s,t} \Big)^{\frac{1}{t+1}}
 = \Big( (P^{0,t})^2 P^{0,r} P^{r,t} \prod_{0<s<t; s\neq r} P^{0,s} P^{s,t} \Big)^{\frac{1}{t+1}}\\
\neq & \hat{P}_{GEKS}^{0,r} \hat{P}_{GEKS}^{r,t} =
\Big( (P^{0,r})^2 \prod_{0<s<r} P^{0,s} P^{s,r} \Big)^{\frac{1}{r+1}}
\Big( (P^{r,t})^2 \prod_{r<s<t} P^{r,s} P^{s,t} \Big)^{\frac{1}{t-r+1}},
\end{align*}
where the set of reference periods is $R(0,t) = \{ 0, 1, ..., t\}$ for $\hat{P}_{GEKS}^{0,t}$, it is $R(0,r) = \{ 0, 1, ..., r\}$ for $\hat{P}_{GEKS}^{0,r}$ and $R(r,t) = \{ r, r+1, ..., t\}$ for $\hat{P}_{GEKS}^{r,t}$, according to the default choice of $R_M$. For example, for $t=2$, the three GEKS indices are intransitive, where
\[
\hat{P}_{GEKS}^{0,2} = \left( (P^{0,2})^2 P^{0,1} P^{1,2} \right)^{\frac{1}{3}}, \quad
\hat{P}_{GEKS}^{0,1} = \left( (P^{0,1})^2 \right)^{\frac{1}{2}} = P^{0,1}, \quad
\hat{P}_{GEKS}^{1,2} = \left( (P^{1,2})^2 \right)^{\frac{1}{2}} = P^{1,2}.
\]
It is of course possible to use a different $R_M$, such as a 13-month window that moves with $t$. But the GEKS remains intransitive, because $P^{0,r}$ calculated with $R_M = \{ r-12, r-11, ...., r\}$ is different from $P^{0,r}$ calculated with $R_M = \{ t-12, t-11, ..., t\}$, provided $t-12\leq 0<r<t$.  
The components in \eqref{GEKS} can be the bilateral MGK, GUV or WGM index, since they are all time reversible. Now that the reference universe of the GEKS index by definition cannot be $U_{R_B}$ for $t > 1$, it  generally does not pass any other tests than the responsiveness test T5.

\section{A reference-quantity-price index family} \label{outline}

None of the indices considered in Section \ref{index} satisfies all the five tests proposed in this paper. Two observations seem worth noting. First, a multilateral index generally does not satisfy the identity test T1 nor the fixed-basket test T2. However, we do not therefore conclude that a bilateral index is preferable to a multilateral index in practice, since none of them is perfect and it is possible to compensate for a small shortcoming in one respect with better properties in others. Second, there is a tension between the bound tests t3 and t4 on the one hand, and the responsiveness test t5 on the other hand. As a potential means to a resolution, we outline below a large index family, which includes the GUV index family as a subclass. Let
\begin{equation}
P_{RQP}^{0,t} =  \Big( P_{RQ}^{0,t} \Big)^{1-\alpha} \Big( P_{GUV}^{0,t} \Big)^{\alpha}, \label{RQP} 
\end{equation}
where $\alpha$ is a constant of choice, for $0\leq \alpha \leq 1$, and $P_{GUV}^{0,t}$ is given by \eqref{GUV}, and the \emph{reference-quantity} index $P_{RQ}^{0,t}$ is given by
\begin{equation} 
P_{RQ}^{0,t} = \sum_{i\in U_{0\cup t}} q_i p_i^t/\sum_{i\in U_{0\cup t}} q_i p_i^0, \label{RQ} 
\end{equation}
where $U_{0\cup t} = U_0\cup U_t$, and $q_i$ is a reference-quantity for $i\in U_{0\cup t}$. We shall refer to \eqref{RQP} as the \emph{reference-quantity-price (RQP)} index. It reduces to a GUV index if $\alpha = 1$.

Provided $0 < \alpha < 1$, an RQP index makes use of both reference quantities $q_i$ and reference prices $p_i$. The expression \eqref{RQP} shows it as a weighted geometric mean of two price indices. It can equally be expressed as deflating the expenditure ratio $V^{0,t}$ by a weighted geometric mean of two quantity indices $Q_{RP}^{0,t}$ and $V^{0,t}/P_{RQ}^{0,t}$, i.e. 
\[
P_{RQP}^{0,t}  = V^{0,t} / \Big[ \big( Q_{RP}^{0,t}\big)^{\alpha} \big( V^{0,t}/P_{RQ}^{0,t} \big)^{1-\alpha} \Big] 
\]
In particular, at $\alpha = 0.5$, the RQP index can be considered to generalise the Fisher index, defined in the special case of $U = U_0 = U_t = U_{0\cup t}$ and $R(0,t) = R_B$. That is, 
\[
P_L^{0,t} P_P^{0,t} = P_L^{0,t} V^{0,t}/Q_L^{0,t} = P_P^{0,t} V^{0,t}/Q_P^{0,t},
\] 
where $P_L^{0,t}$ is the Laspeyres price index given as $P_{RQ}^{0,t}$ with $q_i  = q_i^0$, and $P_L^{0,t}$ is the Paasche price index given as $P_{RQ}^{0,t}$ with $q_i  = q_i^t$, and $Q_L^{0,t}$ is the Laspeyres quantity index given as $Q_{RP}^{0,t}$ with $p_i  = p_i^0$, and $Q_P^{0,t}$ is the Paasche quantity index given as $Q_{RP}^{0,t}$ with $p_i  = p_i^t$.
 
There are many possible choices for the reference price in $P_{GUV}^{0,t}$ and the reference quantity in $P_{RQ}^{0,t}$. In the existing though limited studies and practices of $P_{GUV}^{0,t}$, the reference price $p_i$ is usually set to be the unit-value price of item $i$ over the chosen reference universe, i.e. calculated over the periods in which  the item is available. However, in certain situations, one may instead consider using the introductory price or another representative price. When it comes to the reference quantity $q_i$ in $P_{RQ}^{0,t}$, one can obviously extend the various arithmetic and geometric means defined for the fixed universe. Or, one may set $q_i$ to be the ratio between the average expenditure of item $i$ and the reference price $p_i$ calculated for the GUV-counterpart. In particular, we believe it will be necessary to study these questions together with the formation of homogeneous products, which are defined at a level that is between the items identified by (GTIN, outlet) and the elementary aggregate. However, it is beyond the scope of this paper to address these issues. 

The RQ index $P_{RQ}^{0,t}$ satisfies obviously the identity test T1. It satisfies the fixed-basket test T2 provided $R(0,t) = R_B$. It satisfies the responsiveness test T5, as long as $q_i >0$ for $i\in U_{0\cup t} \setminus U_{0t}$. Moreover, it provides a means to resolve the tension between the bound tests t3 and t4 and the responsiveness test t5. To satisfy the test t5 in the setting of the test t3, where $U_{0\cup t} = U_0 \cup U_{t\setminus 0}$, $p_i^0 = p_i^t$ for $i\in U_0$ and $q_i^t > 0$ for $i\in U_{t\setminus 0}$, we require $P_{RQ}^{0,t} < 1$. Since $\sum_{U_0} q_i p_i^t =  \sum_{U_0} q_i p_i^0$ regardless of the choice of $q_i$ for $i\in U_0$, we need $\sum_{i\in U_{t\setminus 0}} q_i p_i^0 > \sum_{i\in U_{t\setminus 0}} q_i p_i^t$, given any choice of $q_i$ for $i\in U_{t\setminus 0}$. This can be achieved by imputing a price $\hat{p}_i^0$, where $\hat{p}_i^0 > p_i^t$ for $i\in U_{t\setminus 0}$. Provided such $\hat{p}_i^0$, the imputed reference-quantity expenditure in period 0 would be higher than the reference-quantity expenditure in period $t$, i.e.
\[
\sum_{U_0} q_i p_i^0  + \sum_{U_{t\setminus 0}} q_i \hat{p}_i^0 > \sum_{U_0} q_i p_i^0  + \sum_{U_{t\setminus 0}} q_i p_i^t 
= \sum_{U_0} q_i p_i^t  + \sum_{U_{t\setminus 0}} q_i p_i^t.
\] 
It follows that the imputed $P_{RQ}^{0,t}$ is less than 1, which satisfies the upper bound test $t3$ and the responsiveness test t5 at the same time. Similarly, the imputed RQ index $P_{RQ}^{0,t}$ satisfies jointly the tests t4 and t5, provided $\hat{p}_i^t > p_i^0$ for $i\in U_{0\setminus t}$. 

Imputation seems a natural remedy for the RQ index because, unlike the MGK, GUV or WGM index, it lacks otherwise a mechanism that accounts for the differing sizes of the item universes $U_0$ and $U_t$. The inclusion of $\sum_{U_{t\setminus 0}} q_i \hat{p}_i^0$ or $\sum_{U_{0\setminus t}} q_i \hat{p}_i^t$ can be considered as a means to incorporate a dynamic basket from the COGI perspective, or to align the utility over time from the COLI perspective. In the setting of test T3, where $p_i^t < p_i^0$ for at least some $i\in U_0$, we have $\sum_{U_0} q_i p_i^t < \sum_{U_0} q_i p_i^0$ regardless of the choice of $q_i$ for $i\in U_0$. It follows that the imputed $P_{RQ}^{0,t}$ satisfies the test T3, provided any $\hat{p}_i^0 \geq p_i^t$ for $i\in U_{t\setminus 0}$, including the choice of $\hat{p}_i^0 = p_i^t$. Similarly, it satisfies the lower bound test T4, provided $\hat{p}_i^t \geq p_i^0$ for $i\in U_{0\setminus t}$, including the choice of $\hat{p}_i^t = p_i^0$. 

The test results of the RQP index can be deduced from those of $P_{GUV}^{0,t}$ and $P_{RQ}^{0,t}$. Thus, given a judicious choice of the imputed $P_{RQ}^{0,t}$, it can potentially satisfy all the five tests.

\section{Concluding remarks} \label{remark}

The proposed set of tests provide a rigorous diagnostic for whether an index can be considered completely appropriate in a dynamic item universe, as well as pointing towards the directions of possible remedies. The RQP index family can potentially satisfy all the tests. It extends the GUV index family that has received much attention in the recent years. But more research is needed regarding the imputation method and the mixing weight $\alpha$. 

We reiterate that failing one or more tests does not in itself make an index unacceptable in practice, because not exactly satisfying a test does not mean that it is not satisfied approximately, and it is possible for an index to compensate for a small shortcoming in one respect with better properties in others. Moreover, the test approach does not directly provide the solutions to the many other choices one necessarily has to make in practice. These include the use of fixed base period vs. moving base and indirect measurement of the short-term price index, the aggregation structure of the CPI including the formation of homogeneous products, the balance between automatic item-matching and manual intervention, the decision between bilateral and multilateral indices in a given CPI sub-universe, etc. For these reasons, we believe it is important in future research to develop sensible \emph{empirical criteria}, regarding when an index based on the unit value price data can be considered acceptable for practical purposes.

\end{document}